\documentclass[fleqn,10pt]{wlscirep}
\usepackage[utf8]{inputenc}
\usepackage[T1]{fontenc}

\usepackage{siunitx} 
\DeclareSIUnit\bar{bar}

\newcommand{\Ca}{\textsuperscript{40}C\lowercase{a}\textsuperscript{+}}
\newcommand{\neuCa}{\textsuperscript{40}C\lowercase{a}}
\newcommand{\Sl}{S\textsubscript{1/2}}

\newcommand{\Dl}{D\textsubscript{3/2}}
\newcommand{\Dh}{D\textsubscript{5/2}}

\title{A fully fiber-integrated ion trap for portable quantum technologies}

\author[1,*]{Xavier Fernandez-Gonzalvo}
\author[1]{Matthias Keller}
\affil[1]{Department of Physics and Astronomy, University of Sussex, Brighton, BN1 9QH, United Kingdom}
\affil[*]{x.fernandez-gonzalvo@sussex.ac.uk}

\begin{abstract}
Trapped ions are a promising platform for the deployment of quantum technologies. However, traditional ion trap experiments tend to be bulky and environment-sensitive due to the use of free-space optics. Here we present a single-ion trap with integrated optical fibers directly embedded within the trap structure, to deliver laser light as well as to collect the ion's fluorescence. This eliminates the need for optical windows. We characterise the system's performance and measure the ion's fluorescence with signal-to-background ratios on the order of 50, which allows us to perform internal state readout measurements with a fidelity over 99\% in \SI{600}{\micro\second}. We test the system's resilience to thermal variations in the range between \SI{22}{\celsius} and \SI{53}{\celsius}, and the system's vibration resilience at \SI{34}{\hertz} and \SI{300}{\hertz} and find no effect on its performance. The combination of compactness and robustness of our fiber-coupled trap makes it well suited for applications in, as well as outside, research laboratory environments, and in particular for highly compact portable quantum technologies, such as portable optical atomic clocks. While our system is designed for trapping \Ca\ ions the fundamental design principles can be applied to other ion species.
\end{abstract}

\begin{document}

\flushbottom
\maketitle

\thispagestyle{empty}

\section*{Introduction}

Trapped ions are a promising candidate for a wide range of quantum technologies. They are intrinsically reproducible systems, exhibiting long coherence and trapping lifetimes, and techniques to prepare, readout and manipulate their internal and external quantum states are well-developed. This makes them highly suitable to be used in quantum information processing \cite{haff2008, Sage2019}, precision spectroscopy \cite{Schmidt2015} and tests of fundamental physics \cite{cair2017, Haffner2018} amongst others. While there has been remarkable progress in the development and miniaturisation of novel ion trapping structures and associated vacuum systems \cite{choo2017, Prestage2016}, the optical systems needed to manipulate and detect the state of the trapped ions are still mainly based on free-space optics. This leaves a compact ion trap surrounded by a large volume of optical components, which are often susceptible to drifts and vibrations, requiring regular realignment, since free-space optics can lead to beam-pointing instability and hence a deterioration of the system's performance. While for laboratory-based research systems this can be acceptable, for operation outside research laboratories this poses a significant barrier. In particular, the susceptibility of the beam steering and detection optics to vibrations, temperature fluctuations and drifts hinders the use of trapped ions in fieldable metrology and sensor systems.

In recent years there has been progress in integrating the fluorescence detection optics into the ion trap structure using optical fibers \cite{vand2010, brad2011, taka2013}. This eliminates the need for large numerical aperture lenses, which are prone to misalignment and drift and allows an easy connection to the photon detector. However, this comes with the disadvantage that the lack of spacial filtering results in a higher sensitivity to light scattered by the trap electrodes or the surrounding structures. Another approach is to use integrated superconducting single photon detectors \cite{toda2021} and single-photon avalanche photodiodes \cite{setz2021}. While these offer great collection efficiencies they are best suited to planar ion traps as opposed to 3-dimensional trapping structures, the latter being preferred for atomic clock applications due to their lower heating rates and higher trapping efficiencies. Furthermore, the requirement to operate at cryogenic temperatures for superconducting devices prohibits their use in highly compact and portable systems. A third approach is to use in-vacuum integrated optics to maximise the collection of ionic fluorescence \cite{stre2011,merr2011,ghad2017}, working in conjunction with out-of-vacuum optical elements. These solutions are well suited to planar ion traps, and are particularly interesting for multi-ion systems, but they still require a windowed vacuum chamber and careful alignment of external optical components.

Progress has also been made in the integration of the delivery optics, using optical waveguides embedded into the substrate of surface ion traps \cite{niff2020, meht2020, day2021, ivor2021}. Here, diffractive couplers are used to focus the beams onto the position of the ion. This leads to mechanically robust and realignment-free systems,
and produces sufficiently small beam waists. However, aligning the input fibres with the embedded waveguides can be difficult, leading to low overall optical transmission efficiencies. Single-wavelength beam delivery using a single mode optical fibre integrated in a surface trap has also been reported \cite{kim2011}, but so far total integration of all the delivery beams as well as fluorescence collection hasn't been shown.


In this article we present a fiber-integrated ion trap structure, eliminating the need for external free-space optics or optical access. An end-cap style ion trap based on \cite{taka2013} has an optical multimode fiber integrated into one of the rf electrodes for fluorescence collection, and uses in-vacuum optical fibers and focusing optics to deliver the required laser light to the ion. This laser delivery structure facilitates the flexible alignment of the individual beam polarisations and angles during trap assembly. The geometrical arrangement of the multimode collection fiber, its close proximity to the ion and the good mode shape provided by the delivery optics allow us to measure the ion's fluorescence with high signal-to-background ratios even without any spatial filtering. We have characterised the system under different temperature and vibration conditions, which allows us to show that its performance is unaffected by changes of the environmental condition. The compact size, robustness and flexibility of this trap design make it well suited for applications in single ion experiments outside the research lab, with a particular emphases in portable optical atomic clocks.

\section*{The \label{sec:calcium}\Ca\ ion and required wavelengths}

Our system is designed for trapping calcium ions. \Ca\ is particularly well suited for applications in portable optical atomic clocks and sensors because all the wavelengths required for ionisation, cooling, repumping, quenching and spectroscopic interrogation of the clock transition are accessible through compact diode lasers. Moreover, all these wavelengths are compatible with fiber optic components, which is essential for the miniaturisation and ruggedisation of the setup.

\begin{figure}[hb]
\centering
\includegraphics[width=0.55\linewidth]{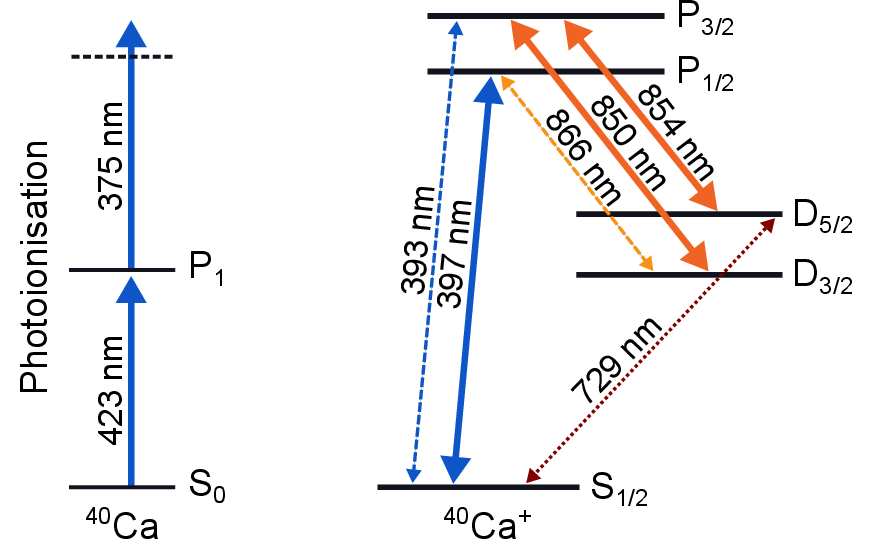}
\caption{Relevant energy levels for the ionisation of \neuCa\ and operation of a \Ca\ atomic clock. In this work we cool the ion using the \SI{397}{\nano\metre} transition, together with \SI{850}{\nano\metre} and \SI{854}{\nano\metre} repumpers. The clock transition in \Ca\ is at \SI{729}{\nano\metre}. Wavelengths have been grouped by colors (blue, orange or red) to represent the beams that can travel through the same type of optical fiber. Solid arrows denote the wavelengths used in this work.}
\label{fig:Ca_levels}
\end{figure}

The relevant energy levels of \neuCa\ and \Ca\ are shown in Fig.~\ref{fig:Ca_levels}. In order to ionise neutral \neuCa\ we use a resonant transition at \SI{423}{\nano\metre} and non-resonant light at \SI{375}{\nano\metre}. We use the cooling transition in \Ca\ at \SI{397}{\nano\metre}, and repumping can be done with \SI{866}{\nano\metre} light or a combination of \SI{850}{\nano\metre} and \SI{854}{\nano\metre} light. \Ca\ has a clock transition at \SI{729}{\nano\metre}. The \SI{854}{\nano\metre} transition can also be used to quench the ion out of the \Dh\ state following the clock interrogation readout step.

\section*{The fibre-integrated ion trap}

\subsection*{Trap geometry}
\begin{figure}[bt]
\centering
\includegraphics[width=0.92\linewidth]{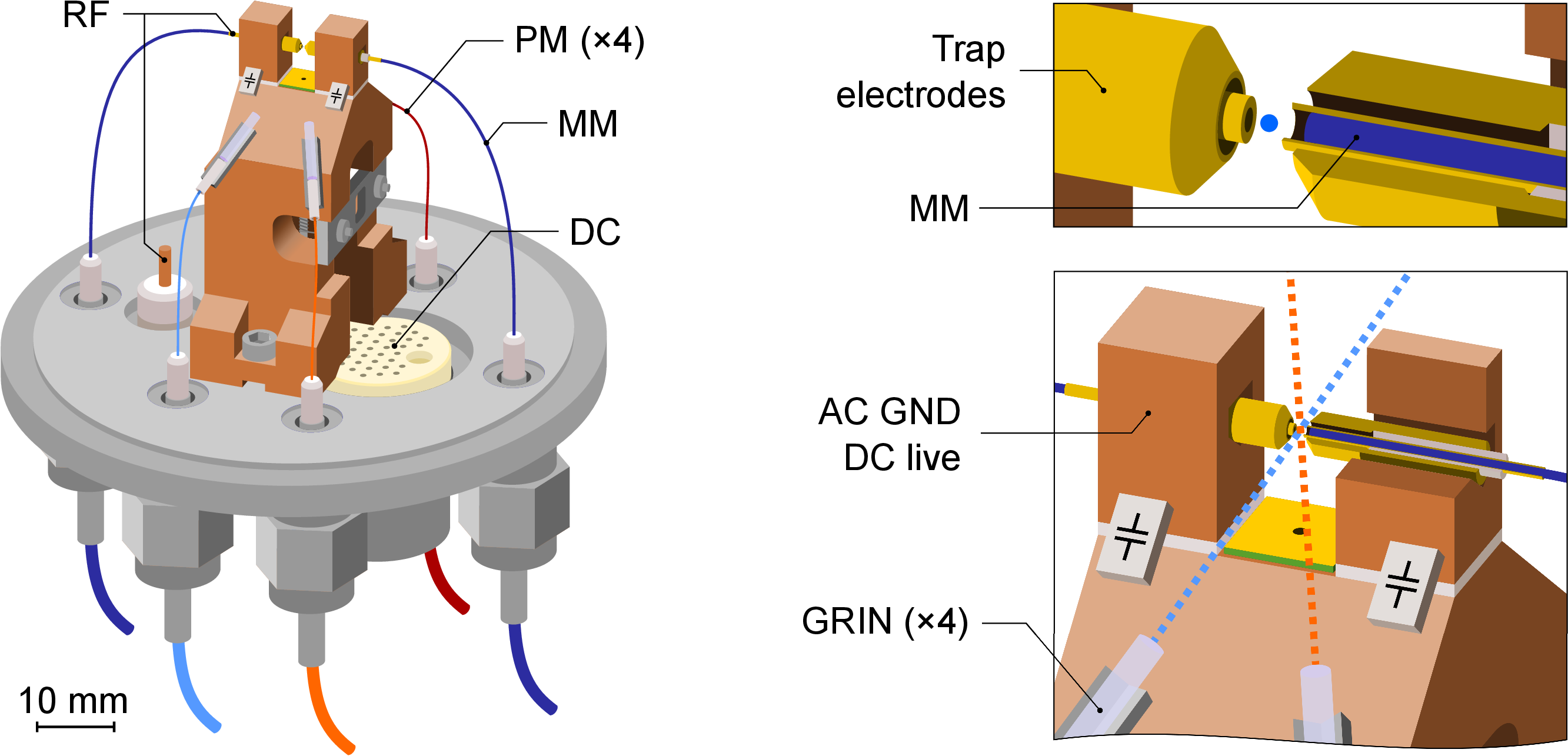}
\caption{Schematic representation of the fiber-integrated ion trap. Temperature sensors, wiring and dc electrodes have been omitted. \emph{Left:} overview of the trap showing the polarisation maintaining (PM) fibers used for light delivery and the multi-mode (MM) fiber used for fluorescence collection, as well as the fiber, dc and rf feedthroughs. \emph{Bottom right:} zoom-in on the trapping structure showing the gradient-index (GRIN) lens collimators and the path followed by the delivery beams, as well as the rf decoupling capacitors. \emph{Top right:} zoom-in and cross-section of the electrode structure, showing the MM fiber embedded inside the rf electrode. The position of the ion is represented with a light blue circle (not to scale).}
\label{fig:trap}
\end{figure}

The trap, schematically shown in Fig.~\ref{fig:trap}, is an end-cap style trap, which provides three-dimensional rf confinement. It consists of two sets of cylindrical concentric electrodes facing each other, with the trap center being in the gap between the electrode assemblies. The inner electrodes are connected to the rf potential, while the outer ones are connected to ground. The inner rf electrodes are hollow, and house multimode fibers, which are used for fluorescence collection. The inner electrodes' outer diameter is \SI{500}{\micro\metre} and they protrude by \SI{250}{\micro\metre} from the ground electrodes. The outer electrodes' inner and outer diameters are \SI{800}{\micro\metre} and \SI{1.78}{\milli\metre} respectively, and they are tapered at \SI{45}{\degree} to increase the optical access angle and prevent clipping the laser beams. An alumina tube is used between the inner and the outer electrodes to electrically isolate them while maintaining concentricity. The electrodes and the alumina spacer are glued together using UVH compatible epoxy (EPO-TEK 353ND).

The axial separation between rf electrodes is \SI{500}{\micro\metre}. The inner electrodes are connected to the main rf source at the back of the electrodes. The outer electrodes are grounded by connecting them to the main body of the trap through a pair of capacitors. This allows them to be used as dc electrodes for micromotion compensation in the axial direction, while keeping them ac grounded. Two dc electrodes are utilised to supply micromotion compensation voltages in the radial plane. A resistively heated tantalum tube filled with calcium is mounted inside the copper body holding the trap, and serves as a calcium dispenser. Two pinholes collimate the calcium atomic beam to pass between the inner electrodes.

\subsection*{Fluorescence collection via multimode fibres}
The integration of the fluorescence collection fiber into the electrode assembly removes the need for alignment, since the fiber is concentric with the rf electrodes and is therefore aligned with the expected position of the ion. The system is therefore insensitive to small misalignments of the fiber position, making it inherently robust to mechanical vibrations and thermal drifts.

The multimode fiber used for fluorescence collection (Thorlabs FG200UEA) has a core diameter of \SI{200}{\micro\metre} and a cladding diameter of \SI{220}{\micro\metre}. The core is made of pure silica, and the cladding is made of fluorine-doped silica. The acrylate protective coating of the fiber was stripped and its end was tapered down to a diameter of \SI{190}{\micro\metre} over \SI{11}{\milli\metre} to provide a tight fit to the rf electrodes' inner bore. The multimode fiber is retracted 90-\SI{100}{\micro\metre} with respect to the rf electrodes' front surface. The fibre is glued at the back of the rf electrode using UHV compatible epoxy (EPO-TEK 301-2).

Light collected in the multimode fiber is spectrally filtered using a narrow band-pass filter and a a photo-multiplier detector (PMT) is then used to measure the ion's fluorescence. Based on the geometry of the system, the fraction of light captured by the fiber is approximately \SI{1.2}{\percent}, limited by its numerical aperture, meaning a total possible of about \SI{2.4}{\percent} if two fibers are used. In this work only one fibre was used, owing to an accidental breakage of the second one during the later stages of the assembly process. Optical losses between the ion and the PMT will comprise: reflection losses at the input and output faces of the MM fibre (3.6\% on each surface, assuming a refractive index of 1.47\cite{mali1965}), propagation losses along the fibre (1\% at \SI{400}{\nano\metre} for a \SI{1}{\metre} fibre) and transmission losses through the band-pass filter (7\% at \SI{397}{\nano\metre}), leading to a total loss of 15\%. With the PMT nominal photon detection efficiency at \SI{400}{\nano\metre} of 30\%, the overall fluorescence detection efficiency is about 0.3\% (0.6\% for both fibers).

\subsection*{Beam delivery via polarisation maintaining fibres}
To deliver the necessary laser beams for the ionisation of \neuCa\ and for the cooling and repumping of \Ca\ ions we use different off-the-shelf optical fibers for different wavelength groups (refer to Fig.~\ref{fig:Ca_levels}). They are all single mode polarisation maintaining fibres. We use an ultraviolet (UV) fiber (Thorlabs PM-S405-XP) to deliver the photoionisation lasers as well as the cooling beam, and a single infrared (IR) fiber (Thorlabs PM780-HP) to deliver the repumper beams at \SI{850}{\nano\metre} and \SI{854}{\nano\metre}. This IR fiber can also be used to deliver light at \SI{866}{\nano\metre}. Additionally, the system is equipped with a second UV fiber for another cooling beam (not used in this work), and a dedicated fiber (Thorlabs PM630-HP) for the future clock laser. The fibers are fed into the vacuum system using optical fiber feedthroughs described in \cite{peer2020}, which were all independently tested to have a leak rate below our measurement limit of \SI{1e-9}{\milli\bar\cdot\litre/\second}.

Anti-reflection-coated gradient-index (GRIN) lenses with a design focal length of \SI{10}{\milli\metre} are employed to focus the fiber outputs into the center of the trap. The delivery fibers sit in a ceramic ferrule just behind the GRIN lenses, with a fiber-to-lens separation of less than \SI{100}{\micro\metre}. These laser delivery systems create close-to-diffraction-limited beams, with a measured beam waist $w_0$ ($1/e^2$ radius) of \SI{5.71(6)}{\micro\metre} and \SI{5.43(2)}{\micro\metre} for the \SI{397}{\nano\metre} beams, \SI{9.82(7)}{\micro\metre} for the \SI{729}{\nano\metre} beams and \SI{11.1(1)}{\micro\metre} for the \SI{866}{\nano\metre} beams. As discussed below, we don't fully exploit the small beam waists, but the good mode shape and absence of beam halos minimises background counts due to the beams scattering on the electrodes. As it will be shown later, this enables us to measure the ion's fluorescence through the multimode fibre with high signal-to-background ratios without any spatial filtering. Note that the \SI{729}{\nano\metre} beam is not used in this work, since probing the clock transition is out of the scope of this initial investigation.

The beams are aligned to the geometrical center of the rf trap during assembly with a combination of a scattering screen placed between the inner electrodes and a pair of microscopes used to observe the laser beam positions. The alignment of the delivery assemblies was done using micro-positioning stages in three dimensions, and we estimate we were able to position the beam within \SI{5}{\micro\metre} from the geometrical center of the trap. In order to increase the robustness against misalignment, the beam foci were positioned such that the beam ($1/e^2$) radius was around \SI{25}{\micro\metre} at the expected position of the ion. Once the alignment was optimised the lenses were glued to the main body of the trap using UHV compatible epoxy (EPO-TEK H21D). The epoxy was cured at \SI{80}{\celsius} for at least 4 hours, during which we manually fed back on the translation stages to keep the beams aligned. After the curing process the beams typically stayed aligned to the centre of the trap to within \SI{10}{\micro\metre}. We attribute the small alignment changes to stress accumulated in the epoxy during the curing process.

\section*{Trap performance}

\begin{figure}[b!]
\centering
\includegraphics[width=0.54\linewidth]{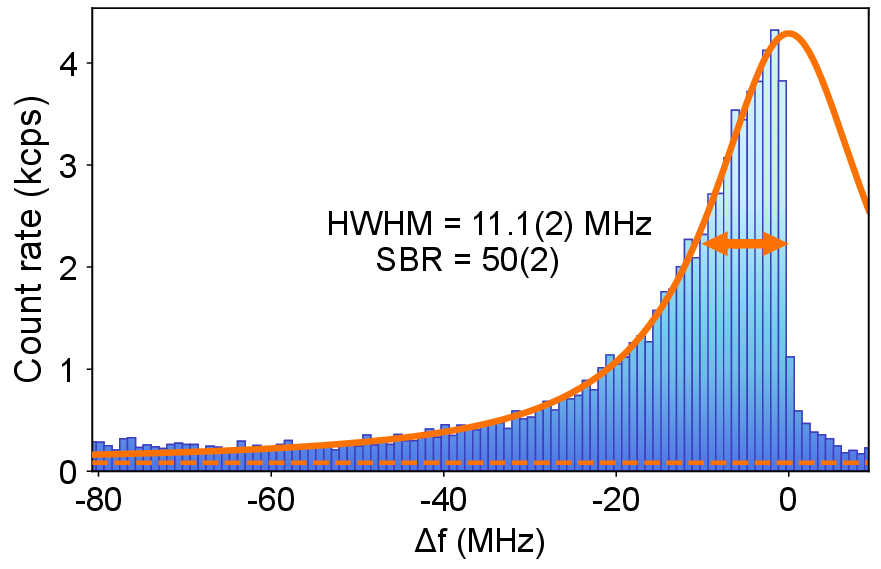}
\caption{Unsaturated cooling transition spectral profile measured at \SI{0.14}{\micro\watt}. The orange solid line is a Lorentzian fit to the red-detuned data, showing a fitted linewidth close to the natural linewidth of the \Ca\ cooling transition. The dashed orange line shows the count rate measured without an ion in the trap, i.e. the background scattering count rate.}
\label{fig:fluorescence_scan}
\end{figure}

To characterise the trap we use a vacuum chamber with an optical window. This allows us to use an sCMOS camera (Andor Zyla) to observe the ion during characterisation, but this is not required to operate the trap.

The system was pumped down to $\lesssim$\SI{ e-10}{\milli\bar} using a getter-ion combination pump (Saes NEXTorr D 100-5). After bakeout and pumping, ions were trapped within the first two days of trying, since no optical alignment was necessary. Both atomic and ionic fluorescence could be observed through the multimode fiber by using an appropriate band-pass filter in front of the PMT.

The trap is driven at a frequency of \SI{13.7}{\mega\hertz} via a resonant transformer. The secular frequencies are kept between \SI{0.6}{\mega\hertz} and \SI{4.5}{\mega\hertz} in the axial direction and between \SI{0.4}{\mega\hertz} and \SI{2.0}{\mega\hertz} in the radial directions. Assuming the trap's $a$-values to be negligible ($a_{x,y,z}\approx0$), the $q$-values are within the ranges $q_{x,y} =$ 0.08 - 0.41 and $q_z =$ 0.12 - 0.92. 

Excess micromotion due to external stray fields is compensated using a combination of the trap depth modulation method and the photon correlation method \cite{berk1998}. From loading to loading, the micromotion compensation voltage values only change by small amounts ($\lesssim$\SI{5}{\percent}) and are otherwise stable.

In contrast to the expected ion lifetime of hours, the ion lifetime within this trap is around 10 minutes. We attribute this to a virtual leak inside the electrode structure. Using a UHV compatible epoxy with a higher viscosity (e.g. EPO-TEK H21D) could have reduce the probability of gas pockets forming between the MM fibres and the rf electrodes due to capillary action.

\subsection*{Cooling transition spectroscopy, signal to background ratio and beam-ion positioning}
With micromotion compensated, we measured the cooling transition spectral profile. These measurements are performed by scanning the frequency of the \SI{397}{\nano\metre} laser using an acousto-optic modulator while recording the fluorescence PMT counts at the output of the multimode fiber. Fig.~\ref{fig:fluorescence_scan} shows a spectrum for a cooling laser power of \SI{0.14}{\micro\watt}. Fitting a Lorentzian function to the data we can extract a half width at half maximum (HWHM) of \SI{11.1(2)}{\mega\hertz} (with the natural transition HWHM being \SI{10.8}{\mega\hertz} \cite{hett2015}). Repeating this measurement for different powers shows that the main contribution to the line broadening is power broadening, with the HWHM at zero power converging to the natural half-width. The signal to background ratio $SBR = (S-B)/B$ (where $S$ is the count rate at the peak of the transition and $B$ is the background count rate measured without an ion) will depend on the cooling laser power due to power broadening. The best values were obtained for powers under \SI{0.2}{\micro\watt}, where power broadening is negligible, with a $SBR$ on the order of 50. For the typical cooling powers used to operate the trap (between 3 and \SI{4}{\micro\watt}), the $SBR$ is on the order of 10 to 20.

Furthermore, we used a series of HWHM measurements at different laser powers to estimate the position of the cooling laser beam with respect to the ion. The laser intensity at the position of the ion can be inferred from power broadening. Comparing this with the actual laser power and the beam waist at the position of the ion, we can calculate where the ion is located within the Gaussian profile of the beam. The ion-to-beam-centre distance was found to be 10.8(1.1)\SI{}{\micro\metre}, with the uncertainty being dominated by the measurement of the laser power at the position of the ion. With a beam waist of \SI{25}{\micro\metre} the ion is well within the cooling laser beam.

\subsection*{State detection fidelity}
\begin{figure}[t!]
\centering
\includegraphics[width=0.52\linewidth]{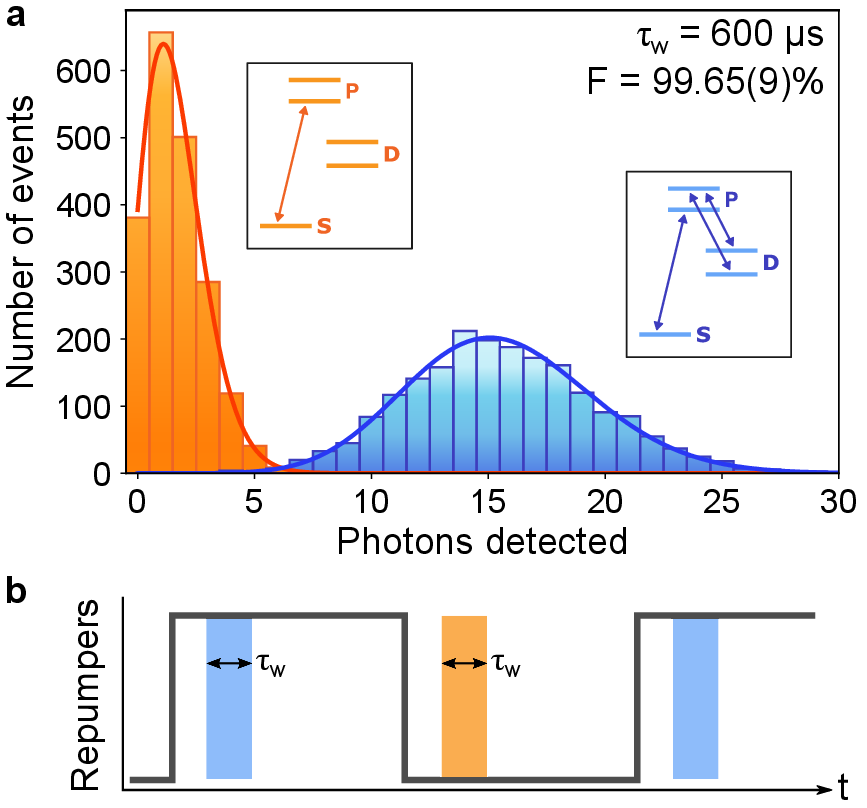}
\caption{\label{fig:state_detection}\textbf{(a)} State detection measurement for a measurement window $\tau_\textrm{w}$ = \SI{600}{\micro\second}. The orange (blue) histogram corresponds to an ion prepared in the dark (bright) state. The lines are Poisson fits to the data, for reference only. \textbf{(b)} Pulse sequence utilised for the state detection measurements. The cooling laser is always kept on, while the repumpers are switched on and off periodically to toggle the ion between the dark and bright states. The shaded areas represent the measurement window time during which counts are added to the bright and dark histograms. There is a \SI{100}{\micro\second} delay between switching the repumpers off (on) and the measurement window, to ensure the ion has been shelved (de-shelved).}
\end{figure}

Next we characterise the state detection fidelity in the trap by preparing the ion in either a bright or a dark state, and comparing the photon counting statistics measured with the PMT. A bright state is obtained by keeping the ion in its cooling cycle, i.e. by keeping the cooling laser on, as well as the repumpers. A dark state is obtained by switching off the repumpers, shelving the ion into the D-states. In terms of determining the state readout fidelity this is equivalent to preparing the ion in either the \Sl\ (bright) or the \Dh\ (dark) state (replicating the shelving that will occur during clock interrogation of the \SI{729}{\nano\metre} transition). The measurement sequence can be seen in Fig.~\ref{fig:state_detection}(b). Photons arriving to the PMT are counted for a time window of length $\tau_\textrm{w}$ for both a dark and a bright ion. The measurements are repeated multiple times, and two histograms are obtained. An example of these can be seen in Fig.~\ref{fig:state_detection}(a).

In order to determine the state of an ion a threshold value $n_{th}$ is defined (along the horizontal axis in Fig.~\ref{fig:state_detection}(a)), above which the ion will be considered to be bright, and below which the ion will be considered to be dark. For the bright state, the detection fidelity is given by:
\begin{equation}
F_B = \frac{\sum\limits_{n=n_{th}}^{\infty}h_B(n)}{\sum\limits_{n=n_{th}}^{\infty}\left[ h_B(n) +h_D(n) \right]},
\end{equation}
with $h_{B,D}(n)$ being the bright and dark histograms as a function of the photon number $n$. Similarly, the detection fidelity for the dark state is given by:
\begin{equation}
F_D = \frac{\sum\limits_{n=0}^{n_{th}-1}h_D(n)}{\sum\limits_{n=0}^{n_{th}}\left[ h_B(n) +h_D(n) \right]}.
\end{equation}
The state detection fidelity is then calculated as the average between the two, $F=\frac{1}{2}\left( F_B+F_D \right)$.

The optimal $n_{th}$ value depends on the detection window time, the cooling and repumper laser powers and their detunings with respect to the line centers. We measured the state detection fidelity for a range of detection window times and cooling laser powers, and we can achieve state detection fidelities better than 99\% for detection periods as short as \SI{600}{\micro\second} (example in Fig.~\ref{fig:state_detection}). The state detection fidelities are calculated directly from the measured data, without correcting for finite state preparation fidelity, finite state lifetime or any other detrimental effects \cite{burr2010}, and that we haven't made any assumptions about the statistical distribution of the measured histograms. Due to low clipping on the electrodes, the low PMT sensitivity to near-infrared light and the band-pass filter, there is no measurable scatter from the repump lasers.

\begin{figure}[t!]
\centering
\includegraphics[width=0.54\linewidth]{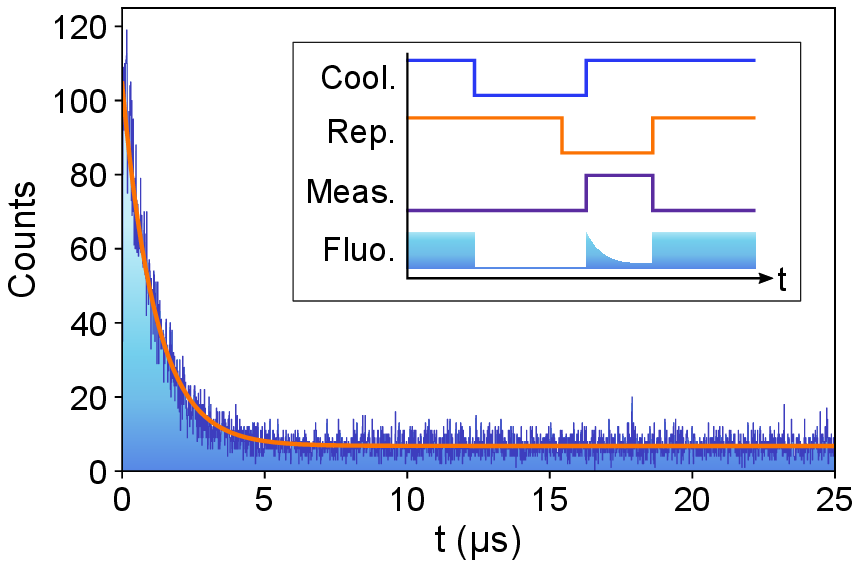}
\caption{Fluorescence decay rate measurement at room temperature for a cooling laser power of \SI{1.6}{\micro\watt}. The orange line is an exponential fit to the data, from which a decay time constant $\tau_\Omega$ can be extracted. \emph{Inset:} pulse sequence used for the measurement of $\tau_\Omega$.}
\label{fig:tau_and_sequence}
\end{figure}

\subsection*{Thermal stability}
The stability of the fiber-integrated ion trap against temperature changes is an important factor for its use outside of research laboratory environments. In order to test the effects of changing temperatures in our trap we measure the optical pumping time $\tau_\Omega$ to the D-states (which is directly related to the laser intensity at the position of the ion) while raising the temperature of the trap. To do so we heat up the entire vacuum chamber using a resistive heating belt, and let the system thermalise for a few minutes. The temperature is measured using three PT100 temperature sensors mounted at different places directly on the trap structure (one on each block holding the electrodes and one on the main copper mount).

\begin{figure}[h!]
\centering
\includegraphics[width=0.55\linewidth]{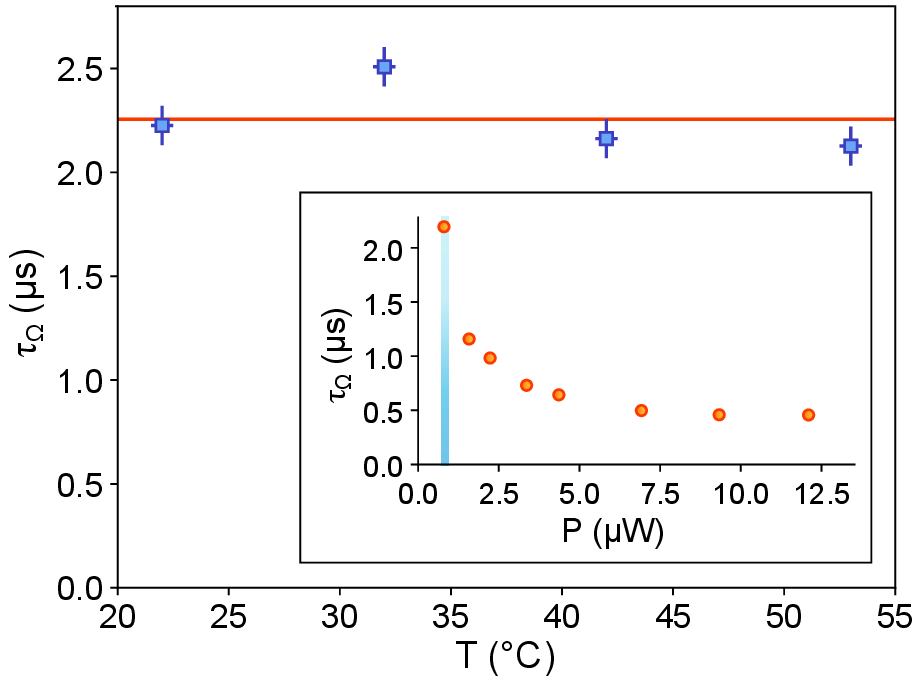}
\caption{Fluorescence decay constant as a function of the trap temperature. The horizontal orange line is the averaged $\tau_\Omega$ between all the measurements. The horizontal error bars represent the statistical error on reading the temperature using the three different thermal sensors. The vertical error bars combine the statistical error in the fit for $\tau_\Omega$ and the error in determining the laser power $P$ times the slope of the $\tau_\Omega$ vs. $P$ curve. \emph{Inset:} fluorescence decay constant as a function of laser power measured at \SI{22}{\celsius}. The shaded blue area denotes the power range at which the data in the main figure were taken.}
\label{fig:tau_vs_P_and_T}
\end{figure}

In order to measure $\tau_\Omega$ we start by preparing the ion in the \Sl\ state, and then switch on the cooling beam with the repumpers off. Fluorescence will be observed until the ion is shelved onto the \Dl\ or the \Dh\ state. Over many repetitions an exponential decay of the fluorescence will be observed (see Fig.~\ref{fig:tau_and_sequence}). The time constant of this decay is $\tau_\Omega$, which is directly related to the Rabi frequency of the cooling beam \cite{Keller2017}. If the beam is misaligned the ion will be exposed to a different light intensity, which in turn will result in a different time constant $\tau_\Omega$. The inset in Fig.~\ref{fig:tau_vs_P_and_T} shows the dependence of $\tau_\Omega$ with the cooling beam power. In order to have high alignment sensitivity to the temperature dependence, measurements were taken using a cooling powers around \SI{0.83(5)}{\micro\watt}, avoiding saturation of the cooling transition while still having an acceptable count rate on the PMT. Fig.~\ref{fig:tau_vs_P_and_T} shows the measured $\tau_\Omega$ for a range of temperatures between \SI{22}{\celsius} and \SI{53}{\celsius}. The variation with respect to the average is consistent with changes in laser power between (and during) the different measurements, which is the main contribution to the uncertainty of these measurements. With the beam centre \SI{10.8}{\micro\metre} away from the position of the ion, a beam waist of \SI{25}{\micro\metre} and a slope of at least \SI{1.3}{\micro\second/\micro\watt} in the blue highlighted section of the inset plot in Fig.~\ref{fig:tau_vs_P_and_T}, and assuming the optical power level to be perfectly stable, the shift in the beam's position is less than $\pm$\SI{1}{\micro\metre}. This is an upper bound, and the actual shift is expected to be much lower, since the variation in $\tau_\Omega$ is fully consistent with the observed variation in laser power (on the order of 5\%). This suggests that thermal expansion and contraction has a negligible effect on beam alignment within the range of temperatures explored.

Another potential issue with changing temperatures is a change in the excess micromotion of the ion, caused by a changing geometry of the trap as it thermally expands or contracts. The micromotion compensation voltages were found to remain constant within 3\% of the average value for all tested temperatures, compatible with the variation observed between different trap loading runs.

\subsection*{Vibration resilience}
Finally, we test the resilience to mechanical vibrations of the fiber-coupled ion trap. To do so, we attach two different sources of vibrations to the vacuum chamber containing the trap, and evaluate its performance. The first vibrating device generates vibrations at frequencies around \SI{34}{Hz} and the second device at around \SI{300}{Hz}. The sCMOS camera looking at the ion is mounted on a floated optical table, in a stationary frame. The vacuum chamber rests on the same optical table but, in order to keep it mechanically isolated, it is loosely clamped to the bench. The result is a system where the vacuum chamber and its contents vibrate but the camera doesn't.

\begin{figure}[h!]
\centering
\includegraphics[width=0.6\linewidth]{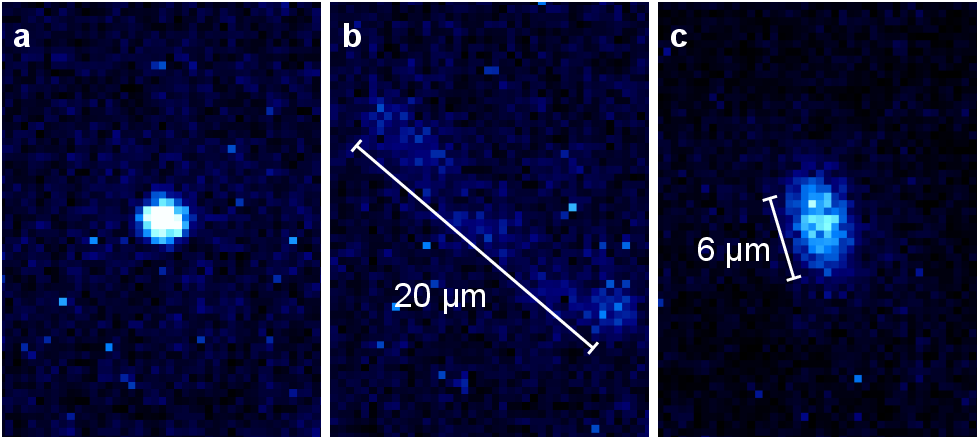}
\caption{\label{fig:vibrations}Camera image comparison between having the vacuum chamber (a) at rest, (b) vibrating at \SI{34}{Hz} with an apparent peak acceleration of 0.047(5)\,g and (c) vibrating at \SI{300}{Hz} with an apparent peak acceleration of 1.09(18)\,g.}
\end{figure}

From the camera images (see Fig.~\ref{fig:vibrations}), assuming the motion of the ion trap to be sinusoidal, the apparent peak accelerations can be calculated for each vibration device. These represent a lower bound to the actual peak acceleration felt by the trap, since the sCMOS camera is only able to capture the ion's motion on a two dimensional plane. When using the first device at \SI{34}{Hz} the apparent peak acceleration is 0.047(5)\,g. For the second device running at \SI{300}{Hz} the apparent peak acceleration is 1.09(18)\,g. In either case, no significant difference is observed either in the ion's fluorescence rate, the micromotion compensation voltages, the cooling transition spectroscopic profile or the fluorescence decay constant $\tau_\Omega$. 

In order to determine the displacement sensitivity to vibrations, we use the atomic ion’s fluorescence. Because we cannot detect any change in the fluorescence level between the situations with and without vibrations, we assume that the fluorescence change due to vibrations is below 10\% of the observed variations due laser power fluctuations. By analysing how a sinusoidal oscillation of the ion’s position with respect to the laser beam influences the ion’s average fluorescence level, we can derive an upper limit of the misalignment amplitude of \SI{3.5}{\micro\metre}. However, we expect the actual amplitude to be considerably smaller.

\section*{Conclusions}

In conclusion, we have presented a compact, fully fiber-integrated, single-ion trap, where optical fibers inside the vacuum chamber are used for beam delivery as well as ion fluorescence collection. The delivery beams are focused onto the expected position of the ion during assembly using GRIN lenses monolithically attached to the trap's body. This makes the system robust against mechanical vibrations and thermal variations, and completely eliminates the need for beam realignment over time. The multimode collection fibers are housed directly inside the trap electrodes, allowing them to sit close to the ion, therefore granting a good solid angle capture and allowing us to measure the ion's fluorescence with high signal to background ratios. We have performed a basic characterisation of the ion trap, including state detection fidelity measurements, and we have subjected the system to a range of temperatures and mechanical vibration conditions, showing no deterioration of its performance.

We believe this is a step forward towards miniaturisation of ion traps for their use in compact and robust integrated systems for applications outside the research laboratory, and specifically for their use in portable optical atomic clocks. Finally, while we use \Ca\ as our ion of choice, the design principles presented here can be extended to other species by choosing fibers and lenses appropriate to the required laser wavelengths.

\bibliography{references}

\section*{Acknowledgements}

This research has received funding from EURAMET (EMPR SIB04-REG4) and the Engineering and Physical Sciences Research Council Quantum Technology Hub for Sensors and Metrology (EP/M013294/1).

\section*{Author contributions statement}

X.F.G. and M.K. conceived the experiment and  X.F.G. conducted the experiment and analysed the results. All authors reviewed the manuscript. 

\section*{Ethics declarations}

\subsection*{Competing interests}
The authors declare no competing interests. 

\section*{Availability of data and materials}
The datasets used and/or analysed during the current study are available from the corresponding author on reasonable request.

\end{document}